\documentstyle[12pt,epsfig]{article}
\begin{document}
\thispagestyle{empty}
\def\ltap{\ \raisebox{-.4ex}{\rlap{$\sim$}} \raisebox{.4ex}{$<$}\ }
\def\gtap{\ \raisebox{-.4ex}{\rlap{$\sim$}} \raisebox{.4ex}{$>$}\ }
\setcounter{page}0
\rightline{Ref. SISSA 5/2000/EP}
\rightline{February 2000}
\vskip 0.6cm
\begin{center}
{\bf On the New Conditions for a Total Neutrino 
Conversion in a Medium
} 

\vspace{0.3cm} 
M. V. Chizhov

\vspace{0.3cm}
{\em
Centre for Space Research and Technologies, Faculty of Physics,\\
University of Sofia, 1164
Sofia, Bulgaria\\
E-mail: $mih@phys.uni$-$sofia.bg$
}

\vspace{0.4cm}
S. T. Petcov \footnote{Also at: Institute of Nuclear Research and
Nuclear Energy, Bulgarian Academy of Sciences, 1784 Sofia, Bulgaria}

\vspace{0.3cm}   
{\em
Scuola Internazionale Superiore di Studi Avanzati, I-34014 Trieste,
Italy, and\\
Istituto Nazionale di Fizica Nucleare, Sezione di Trieste, I-34014
Trieste, Italy
}
\end{center}
\vskip 0.4cm
\begin{abstract}
We show that the arguments  
forming the basis for the claim 
that  the conditions for 
total neutrino conversion derived and studied 
in detail in \cite{ChPet991,ChPet992}
``are just the conditions of the parametric 
resonance of neutrino oscillations
supplemented by the requirement 
that the parametric enhancement
be complete'', given in \cite{ASNPB}
have flaws which make the claim 
physically questionable.  
We show also that in the case of
the transitions in the Earth of 
the Earth-core-crossing
solar and atmospheric neutrinos
the peaks in the relevant transitions probabilities
$P_{a b}$, associated with the new conditions,
$max~P_{a b} = 1$,
are of physical relevance - in contrast
to what is suggested in \cite{ASNPB}.
Actually, the enhancement
of $P_{a b}$ in any region of the 
corresponding parameter
space are essentially 
determined by these absolute maxima 
of $P_{a b}$.
We comment on few other
aspects of the results derived 
in \cite{ChPet991,ChPet992,SP98}
which have been misunderstood and/or
misinterpreted in \cite{ASNPB}. 

\end{abstract}

\vfill
\newpage

\indent 
  
1.  In a {\it Comment} on our results derived and discussed briefly 
in \cite{ChPet991} and in detail
in \cite{ChPet992} 
and on the results derived by one of us in 
\cite{SP98}, Akhmedov and Smirnov
claim \cite{ASNPB}, in particular, 
that ``the conditions for total neutrino conversion studied 
by Chizhov and Petcov \cite{ChPet991}
\footnote{By using the verb ``studied'' 
in connection with our new conditions for a 
total neutrino conversion in a medium 
\cite{ChPet991,ChPet992}, the authors 
of \cite{ASNPB} suggest indirectly that 
these conditions were already
considered in the literature 
before the publication of our papers. 
We would like to note that
none of our {\it new conditions 
for a total neutrino conversion in a medium}
were derived, postulated and/or 
discussed in some form
in the literature on the subject
published before the articles \cite{ChPet991,ChPet992}.}
are just the conditions of the parametric 
resonance of neutrino oscillations
supplemented by the requirement 
that the parametric enhancement
be complete.'' 
We show below that the arguments  
forming the basis for these claims
have flaws which, in our opinion, 
make the claims physically questionable. 
We show also that in the case of
the transitions in the Earth of 
the Earth-core-crossing
solar and atmospheric neutrinos
the peaks in the relevant transitions probabilities
$P_{\alpha \beta}$, associated with the new conditions,
$max~P_{\alpha \beta} = 1$,
are of physical relevance - in contrast
to what is suggested in \cite{ASNPB}.
Actually, the enhancement
of $P_{\alpha \beta}$ in any region of the 
corresponding parameter
space is essentially 
determined by these absolute maxima 
of $P_{\alpha \beta}$.
We comment on few other
aspects of the results derived 
in \cite{ChPet991,ChPet992,SP98}
which have been misunderstood and/or
misinterpreted in \cite{ASNPB}. 

  The form of the {\it new conditions} for a 
total neutrino conversion in a 
{\it medium consisting of two or three 
(nonperiodic) constant density layers},
derived in \cite{ChPet991,ChPet992}, 
the region of the parameter space 
(i.e., the $\Delta m^2/E - \sin^22\theta$ plane) 
where they can be realized, and the 
physical interpretation of the corresponding
absolute maxima of 
the neutrino transition probabilities of interest 
as being caused 
by a {\it maximal constructive interference} 
between the amplitudes of the neutrino transitions
in the (two) different constant density layers,
found in our studies,
made us conclude in \cite{ChPet991,ChPet992} 
that these {\it new conditions}  
differ from 
the conditions for parametric 
resonances in the neutrino transitions,
discussed in the articles 
\cite{Param86,Akh88,KS89}
and possible in a medium with density, 
varying periodically along the neutrino path.
The {\it Comment} \cite{ASNPB} does not
provide viable arguments
against our conclusions. 

2. Consider transitions of neutrinos crossing
a medium consisting of 
i) two layers with different constant
densities $N_{1,2}$ and widths $L_{1,2}$, 
or of ii) three layers of constant
density, with the first and the third layers
having identical densities
$N_{1}$ and widths $L_{1}$,
which  differ from those of the second layer,
$N_{2}$ and $L_{2}$ \cite{SP98,ChPet991,ChPet992}. Suppose the
transitions are caused by
two-neutrino mixing in vacuum 
with mixing angle $\theta$.  
Let us denote by $\theta_{i}$ and 2$\phi_{i}$, $i=1,2$, 
the mixing angle in matter in the layer with density 
$N_{i}$ and the phase difference
acquired by the two neutrino energy-eigenstates
after neutrinos have crossed this layer.
It proves convenient
to introduce the quantities (see, e.g., \cite{ChPet992}):
\begin{equation} 
\cos\Phi \equiv Y = c_1c_2 - 
\cos(2\theta_2 - 2\theta_1)s_1s_2,
\label{1}
\end{equation}
\begin{equation} 
{\bf X}^2 = 1 - Y^2,
\label{2}
\end{equation}
\begin{equation} 
n_3\sin\Phi \equiv X_3 = - (s_1c_2 \cos 2\theta_1 + 
c_1s_2 \cos 2\theta_2),
\label{3}
\end{equation}
\noindent $X_3$ being the third component of the vector 
\footnote{To facilitate the understanding of the main points 
of our criticism of \cite{ASNPB}, 
we use the same notations as
in \cite{ASNPB} for most of the quantities discussed.} 
${\bf X} = (X_1,X_2,X_3)$, whose first two components 
are also given in terms of $\theta_{i}$ and $\phi_{i}$
(see \cite{ChPet992}).
The probability of the
transition $\nu_{a} \rightarrow \nu_{b}$
(i.e., $\nu_{e} \rightarrow \nu_{\mu (\tau)}$,
$\nu_{\mu} \rightarrow \nu_{e}$, etc.)
after neutrinos have crossed $n$ alternating 
layers with densities $N_{1}$ and $N_{2}$ 
is given according to \cite{ASNPB} by:
\begin{equation}  
P(\nu_a \rightarrow \nu_b;~nL) = 
\left(1 - \frac{X_3^2}{{\bf X}^2}\right)\sin^2\Phi_p 
\equiv 
 \frac{X_1^2 + X_2^2}{X_1^2 + X_2^2  + X_3^2}\sin^2 \Phi_p,
\label{PAS}
\end{equation}
\noindent where $\Phi_p = (n/2)\Phi$ if 
the number of layers $n$ is even, and  
\begin{equation} 
\Phi_p = \frac{n-1}{2}\Phi + \varphi~,~~~
\varphi = \arcsin\left(s_1 \sin 
2\theta_1/\sqrt{1-X_3^2/|{\bf X}|^2}\right)\,
\label{phiAS}
\end{equation}
\noindent for odd number of layers with 
the first layer having density $N_1$.

  The first thing to note is that 
the expression for $\varphi$ in eq. (\ref{phiAS}),
given in \cite{ASNPB}, is 
strictly speaking 
incorrect:
it is valid only if $Z \geq 0$, where 
\begin{equation} 
Z = s_2\sin 2\theta_2 + s_1(\sin 2\theta_1)Y.
\label{Z}
\end{equation}
\noindent                             
The correct expression for 
$\varphi$ for arbitrary $sign~Z$ reads:    
\begin{equation}
\varphi = \arctan\left(s_1\sin 
(2\theta_1)~|{\bf X}|/Z\right)~.
\label{phiCP}
\end{equation}
The authors of \cite{ASNPB} demonstrate the same
imprecision in eq. (2) of their {\it Comment}:
the functions $\arccos Y$ and $\arcsin |{\bf X}|$
have different defining regions and it is 
incorrect to write
$\arccos Y = \arcsin |{\bf X}|$:
this equality is obviously wrong 
when $Y < 0$.  

  According to the authors of \cite{ASNPB},
``Eqs. (\ref{PAS}) and (\ref{phiAS}) describe 
the parametric oscillations
with the pre-sine factor in (\ref{PAS}) and $\Phi_p$
being the oscillation depth and phase.'' and further
``Parametric resonance occurs when the depth 
of the oscillations becomes equal to unity. The resonance
condition is therefore $X_3 = 0$.''

 In the two-layer case i) considered by 
us in \cite{ChPet991,ChPet992}, 
which is relevant for the present discussion,
one has  $\Phi_p = \Phi$ ($n =2$).
This result and eqs. (\ref{1}) and (\ref{2}) imply that
actually $\sin^2 \Phi_p = X_1^2 + X_2^2  + X_3^2$.
Correspondingly, the ``parametric-resonance''
form in which the authors of \cite{ASNPB} 
cast the probability $P(\nu_a \rightarrow \nu_b;~nL)$ 
in this case is artificial: the probability 
is given by \cite{ChPet992}
\begin{equation} 
P(\nu_a \rightarrow \nu_b ;~2L) = 
X_{1}^2 + X_{2}^2 = 1 - Y^2 - X_{3}^2,
\label{PCP2L}
\end{equation}
\noindent where we have used eq. (\ref{2}). 
Therefore any resonance interpretation of the probability
$P(\nu_a \rightarrow \nu_b ;~2L)$
based solely on the eq. (\ref{PAS}) 
with $n = 2$ seems to us physically unjustified.
The {\it new conditions for a total 
neutrino conversion in a medium}
follow in this case from the form (\ref{PCP2L})
of $P(\nu_a \rightarrow \nu_b ;~2L)$ and 
read \cite{ChPet991,ChPet992}:
\begin{equation} 
Y = 0,~~~~~ X_3 = 0.
\label{cond2L}
\end{equation}
\noindent It should be clear from
eqs. (\ref{PCP2L}) and (\ref{cond2L})
that the condition $X_{3} = 0$ {\it alone does not 
ensure the existence even of a local maximum
of the probability $P(\nu_a \rightarrow \nu_b ;~2L)$}. 

  It is not guaranteed {\it a priori} that 
the equations in (\ref{cond2L}) have non-trivial 
solutions in general, and in the specific case
of transitions of neutrinos crossing the Earth 
core and the Earth mantle on the way to the detector.
As we have shown in \cite{ChPet991,ChPet992},
the solutions of the conditions (\ref{cond2L}) 
i) exist and are given by 
\begin{equation}
solution~A^{(2)}:~\left\{ \begin{array}{l}
\tan\phi_1=\pm\sqrt{{\displaystyle -\cos(2\theta_2)\over
\displaystyle\cos(2\theta_1)\cos(2\theta_2 - 2\theta_1)}}~, \\
\tan\phi_2=\pm\sqrt{{\displaystyle -\cos(2\theta_1)\over
\displaystyle\cos(2\theta_2)\cos(2\theta_2 - 2\theta_1)}}~,
\end{array} \right.
\label{max-2A}
\end{equation}
\noindent where the signs are correlated, and ii) 
they can be realized {\it only in the region 
determined by the three
inequalities}:
\begin{equation}
region~A^{(2)}:~\left\{ \begin{array}{l}
\cos 2\theta_1 \ge 0 \\
\cos 2\theta_2 \le 0 \\
\cos(2\theta_2 - 2\theta_1) \ge 0.
\end{array} \right.
\label{region2A}
\end{equation}
\noindent 
It was demonstrated in \cite{ChPet991,ChPet992} 
as well that
the two conditions in eq. (\ref{cond2L}), or equivalently the 
solutions expressed by eqs. (\ref{max-2A}) and
(\ref{region2A}), 
are conditions for a {\it maximal constructive 
interference} between the amplitudes of the neutrino 
transitions in the two layers.
Thus, a natural physical interpretation
of the absolute maxima of 
$P(\nu_a \rightarrow \nu_b ;~2L)$
associated with the conditions (\ref{cond2L}) is that 
of {\it constructive interference maxima}.

  It should be clear from 
the above arguments
that we do not see physical reasons 
to call $X_3 = 0$ 
in the case under discussion 
a ``parametric resonance condition''.
Using the trick of \cite{ASNPB} 
one can easily cast 
the probability of two-neutrino 
oscillations in vacuum and in matter 
with constant density, for example, 
in the ``parametric-resonance'' form (\ref{PAS}).
The analog of the  condition $X_3 = 0$
reduces in these cases respectively to 
$\sin^22\theta = 1$ and $\sin^22\theta_m = 1$,
$\theta_m$ being the mixing angle in matter.
Thus, according to the 
terminology suggested
in \cite{ASNPB}, one should call
the condition  $\sin^22\theta = 1$ and the
MSW resonance condition $\sin^22\theta_m = 1$
``parametric-resonance conditions''.
One can use such a terminology, of course,
but this is not justified by the physics 
of the process and nobody uses it. 
The situation in the two-layer case
discussed above is in essence the same.

  Analogous results and conclusions are valid 
in the case of medium with three layers 
(case ii)) considered in \cite{SP98,ChPet991,ChPet992}
($n = 3$ in eqs. (\ref{PAS}) and (\ref{phiAS})).  
Using the correct expression for $\varphi$,
eq. (\ref{phiCP}), one finds again that the 
``parametric resonance'' form in which  
the authors of \cite{ASNPB} 
write the probability of interest
$P(\nu_a \rightarrow \nu_b ;~3L)$
is artificial: the probability has the same
form as in eq. (\ref{PCP2L}) \cite{ChPet992}
\begin{equation} 
P(\nu_a \rightarrow \nu_b ;~3L)  
= 1 - \bar{Y}^2 - \bar{X}_{3}^2,
\label{PCP3L}
\end{equation}
\noindent where 
\begin{equation} 
\bar{Y} = - c_2 + 2c_1Y, 
\label{barY}
\end{equation}
\noindent and 
\begin{equation} 
\bar{X}_{3} = - s_2 \cos 2\theta_2 - 2s_1 \cos (2\theta_1)~Y .
\label{barX3}
\end{equation}
\noindent The conditions for 
a total neutrino conversion in this case
read \cite{ChPet992}:
\begin{equation} 
\bar{Y} = 0,~~~~~ \bar{X}_3 = 0.~~~~~~~
\label{cond3L}
\end{equation}
\noindent Their solutions in the case of 
the inequality
$N_1 < N_2$ (corresponding to the
relation between the densities 
in the Earth mantle and core)
were given in \cite{ChPet991,ChPet992} 
and have the form:
\begin{equation}
solution~A^{(3)}:~\left\{ \begin{array}{l}
\tan\phi_1=\pm\sqrt{{\displaystyle -\cos 2\theta_2\over 
\displaystyle\cos(2\theta_2 - 4\theta_1)}}, \\
\tan\phi_2 =\pm{\displaystyle \cos 2\theta_1\over \sqrt{
\displaystyle-\cos(2\theta_2)\cos(2\theta_2 - 4\theta_1)}},
\end{array} \right.
\label{max-3A}
\end{equation}  
\noindent where the signs are again correlated.
The solutions can only be realized in the region
\begin{equation}
region~A^{(3)}:~\left\{ \begin{array}{l}
\cos(2\theta_2) \le 0, \\ 
\cos(2\theta_2 - 4\theta_1) \ge 0.
\end{array} \right.
\label{region3A}
\end{equation}  

 It is easy to show that for any number of layers $n$,
the denominator in (\ref{PAS}) 
is always canceled by $\sin^2\Phi_p$
and $P(\nu_a \rightarrow \nu_b;~nL)$
is just  a polynomial without
any resonance-like feature. Indeed, for even $n$, 
as can be shown, we have  
\begin{equation}
\sin\Phi_p=\sin\Phi~U_{n/2-1}(\cos\Phi), 
\label{Phipeven}
\end{equation}
\noindent where $U_n(x)$ is
well-known Chebyshev's polynomial of the second kind 
\cite{Rizhik}.
In the case of odd number of layers ($n \geq 3$) 
one finds
\begin{equation}
P(\nu_a \rightarrow \nu_b;~nL) = 
\left[
s_1\sin2\theta_1\cos\left(\frac{n-1}{2}\Phi\right)+
Z U_{\textstyle\frac{n-3}{2}}(\cos\Phi)
\right]^2.
\label{PCheb}
\end{equation}

  Finally, similar considerations apply to the 
probability of the $\nu_2 \rightarrow \nu_e$
transitions, $\nu_2$ being the heavier 
of the two vacuum mass-eigenstate neutrinos, 
in the three-layer case ii),
$P(\nu_2 \rightarrow \nu_e ;~3L)$. This 
probability 
can be used to account for 
the the Earth matter effects
in the transitions of solar neutrinos traversing the Earth:
$P(\nu_2 \rightarrow \nu_e ;~3L)$
corresponds \cite{SP98} to 
the case of solar neutrinos crossing the
Earth mantle, the core and 
the mantle again on the way to the detector.
As was shown in \cite{ChPet992},
the conditions for a total 
$\nu_2 \rightarrow \nu_e$ conversion,
$max~P(\nu_2 \rightarrow \nu_e ;~3L) = 1$, read:
\begin{equation} 
\bar{Y} = 0,~~~~~ \bar{X}'_3 = 0,~~~~~~~
\label{cond2e3L}
\end{equation}
\noindent where $\bar{Y}$ is given by eq. (\ref{barY}) and  
\begin{equation}
\bar{X}'_{3} = - s_2 \cos (2\theta_2 -\theta) - 
2s_1 \cos (2\theta_1 -\theta)~Y.
\label{barX'3}
\end{equation}
\noindent The solutions of the 
conditions (\ref{cond2e3L})
providing the absolute maxima of 
$P(\nu_2 \rightarrow \nu_e ;~3L)$
and the region where these solutions can take place
were given in \cite{ChPet991,ChPet992};
they can formally be obtained from eqs. (\ref{max-3A})
and (\ref{region3A})
by replacing $2\theta_1$ and $2\theta_2$ 
with ($2\theta_1 - \theta$) and ($2\theta_2 - \theta$).
  
   The three sets of two conditions, eqs. 
(\ref{cond2L}), (\ref{cond3L}) and (\ref{cond2e3L}), and/or 
their solutions (e.g., eqs. (\ref{max-2A}) and (\ref{max-3A})), 
and/or the regions where the solutions
can be realized (e.g., eqs. (\ref{region2A}) and (\ref{region3A})),
were not derived and/or discussed in any form  
in \cite{Param86,Akh88,KS89}
or in any other article on the subject
of neutrino transitions in a medium
published before \cite{ChPet991,ChPet992}. 
None of them follows
from the conditions
of enhancement of 
$P(\nu_a \rightarrow \nu_b)$
found in \cite{Param86,Akh88,KS89}
and thus they not a particular case of the latter. 
That is the reason we used the term ``new conditions for 
a total neutrino conversion in a medium'' for them.

  The authors of \cite{ASNPB} write further:
``One well known realization of the parametric
resonance condition'', i.e., of $X_3 = 0$, is  
\cite{Param86,Akh88,KS89,SP98}~
\footnote{We quote here only 
the references which, in our opinion, are
relevant for the present discussion.} 
``$c_1 = c_2 = 0$, or 
\begin{equation}
2\phi_1 = \pi + 2\pi k_1,~~~2\phi_2 = \pi + 2\pi k_2,~k_1,k_2 =0,1,2,..., 
\label{c1c20}
\end{equation}
\noindent independently of the mixing angles.''
Contrary to what the authors of \cite{ASNPB} claim,
the two conditions in eq. (\ref{c1c20})
were not given in the articles 
\cite{Param86,Akh88,KS89}: what
one finds in these articles {\it at most}  
is the condition $2\phi_1 + 2\phi_2 = 2\pi + 2\pi k$ 
which is not equivalent to 
the two conditions in eq. (\ref{c1c20}). 
The two conditions in eq. (\ref{c1c20})
were discussed in detail for the 3-layer case 
in \cite{SP98}.
Moreover, as we have shown in 
\cite{ChPet992} and would like to 
emphasize here again,
{\it the conditions
$c_1 = 0,~c_2 = 0$ by themself do not lead to
a maximum of the neutrino transition
probabilities of interest
in the neutrino energy variable, unless 
a third nontrivial condition is fulfilled}.
This third condition has the following form
for the $\nu_a \rightarrow \nu_b$ 
(i.e., $\nu_e \rightarrow \nu_{\mu (\tau)}$,
$\nu_{\mu} \rightarrow \nu_{e}$, etc.)
transitions in the two-layer (n = 2) and three-layer
(n=3) medium cases i) and ii), respectively \cite{ChPet992}:
$\cos(2\theta_2 - 2\theta_1) = 0$ 
and $\cos(2\theta_2 - 4\theta_1) = 0$.
For the probability
$P(\nu_2 \rightarrow \nu_e ;~3L)$ 
it reads \cite{ChPet992}:
$\cos(2\theta_2 - 4\theta_1 + \theta) = 0$.
It is not difficult to convince oneself
that the indicated sets of {\it three} conditions represent
possible solutions respectively of (\ref{cond2L}),
(\ref{cond3L}) and (\ref{cond2e3L}) \cite{ChPet992}.
More generally, the condition
$X_{3} = 0$ {\it alone  does not guarantee 
the existence even of a local maximum
of the neutrino transition 
probabilities of interest}.

  In what regards the article by
Q.Y. Liu and A. Yu. Smirnov
quoted in \cite{ASNPB} in connection with
the conditions (\ref{c1c20}) (see ref. [5] in \cite{ASNPB}),
these authors noticed that in the case of 
muon neutrinos crossing the Earth along the specific
trajectory characterized by a Nadir angle $h \cong 28.4^{\circ}$, and for
$\sin^22\theta \cong 1$ and
$\Delta m^2 /E \cong (1 - 2)\times 10^{-4}~{\rm eV^2/GeV}$,
the $\nu_{\mu} \rightarrow \nu_s$ transition
probability, $\nu_s$ being a 
sterile neutrino, is enhanced. 
The authors interpreted this enhancement 
as being due to the conditions 
$2\phi_j = \pi$, $j=1,2$, which 
they claimed to be approximately satisfied.   
Actually, for the values of the parameters
of the examples chosen by Q.Y. Liu and A. Yu. Smirnov
to illustrate their conclusion
one has  $2\phi_1 \cong (0.6 - 0.9)\pi$ and 
$2\phi_2 \cong (1.2 - 1.5)\pi$. 
The indicated 
enhancement is due to \cite{ChPet992} the existence
of a {\it nearby total neutrino conversion point} 
which for $h \cong 28.4^{\circ}$ is located
at  $\sin^22\theta \cong 0.94$ and
$\Delta m^2 /E \cong 2.4\times 10^{-4}~{\rm eV^2/GeV}$
and at which $2\phi_1 \cong 0.9\pi$ and 
$2\phi_2 \cong 1.1\pi$. 
We have also found in \cite{ChPet992} that 
for each given $h \ltap 30^{\circ}$
there are several 
total neutrino conversion points at 
large values of $\sin^22\theta$,
at which the phases
$2\phi_1$ and $2\phi_2$ are not necessarily 
equal to $\pi$ or to odd multiples of $\pi$
(see Table 3 in \cite{ChPet992}).
Thus, the explanation of the enhancement
offered by the indicated authors
is at best qualitative and incorrect in essence.

  That $X_{3} = 0$ {\it alone} 
is a condition for {\it local maxima}
of the probability 
$P(\nu_a \rightarrow \nu_b ;~nL)$
was suggested in E. Kh. Akhmedov,
Nucl. Phys. B538 (1999) 25 (to be quoted 
further as NP B538, 25),
on the basis of the form of the probability
in eq. (\ref{PAS}).
However, as we have already emphasized,
the condition
$X_{3} = 0$ alone {\it does not guarantee 
the existence even of a local maximum 
of the neutrino transition probabilities
$P(\nu_a \rightarrow \nu_b ;~2L)$ and
$P(\nu_a \rightarrow \nu_b ;~3L)$
}.
This should be clear
from the ``natural''  expressions 
for the probabilities
$P(\nu_a \rightarrow \nu_b ;~2L)$ and
$P(\nu_a \rightarrow \nu_b ;~3L)$, given by 
eqs. (\ref{PCP2L}) and (\ref{PCP3L}). 
Of the three solutions
for the extrema of 
$P(\nu_a \rightarrow \nu_b ;~3L)$
found in NP B538, 25,
the solution $c_1 = 0,~c_2 = 0$ 
was already discussed 
in detail in ref. \cite{SP98} 
(compare  eqs. (11) - (16) and (24) in \cite{SP98} with
conditions (1) on page 37 of NP B538, 25), while 
the other two correspond to MSW transitions
\footnote {Let us note that
in what regards the cases of neutrino transitions
studied in \cite{SP98,ChPet991,ChPet992}, the
article NP B538, 25,
contains a rather large 
number of incorrect 
statements and conclusions. 
Most of these statements are concentrated in Section 4 of 
NP B538, 25, where the author discusses 
the realistic case of transitions 
of neutrinos crossing the Earth core we were 
primarily interested in in \cite{SP98,ChPet991,ChPet992}.
}
(they were also briefly discussed  
in \cite{SP98}).

4. The authors of \cite{ASNPB} claim that
``The existence of strong enhancement peaks in transition
probability P rather than the condition P=1 is of physical
relevance.'', although they do not give an example 
of a relevant strong enhancement peak (i.e., local maximum). 
As we have shown in \cite{ChPet992} (see also \cite{SP98}),
the solutions given by eqs. (\ref{max-2A}) and 
(\ref{max-3A}) and those for the probability
$P(\nu_2 \rightarrow \nu_e ;~3L)$
are realized in 
the transitions in the Earth of the 
Earth-core-crossing  neutrinos (solar, atmospheric,
accelerator) and lead to observable effects 
in these transitions.   
From the extensive numerical 
studies we have performed
in the realistic cases of
transitions of neutrinos in the Earth (e.g.,
neutrinos crossing the Earth 
core on the way to the detector)
{\it we do not have any evidence about the presence 
of significant local maxima in the neutrino 
transition probabilities of interest not related to the
peaks of total neutrino conversion}.
Actually, our studies show that only the peaks of total 
neutrino conversion are dominating  
in $P(\nu_a \rightarrow \nu_b ;~2L)$,
$P(\nu_a \rightarrow \nu_b ;~3L)$
and $P(\nu_2 \rightarrow \nu_e ;~3L)$,
and correspondingly determine the regions where 
these probabilities
can be significant
in the corresponding space of parameters. 
The peaks considered, e.g., in \cite{SP98}
(see figs. 1 and 2) 
are points on the ``ridges'' formed by local
maxima, e.g., in the energy variable at fixed
values of the other parameters, leading to 
the peaks of total neutrino conversion, 
discovered by us (see figs. 6 - 9  and Tables 5 - 6 in 
\cite{ChPet992}). As one can convince oneself using 
Figs. 6 - 9 from \cite{ChPet992}, all maxima in
Figs. 4, 5, 10 - 13 and 15 in \cite{s5398}, 
including the relatively small local ones, 
are related to (and determined by) the presence of
corresponding points (peaks) of total neutrino conversion. 

 5. The authors of \cite{Param86,KS89} studied
the effects of {\it small} density perturbations
on the neutrino oscillations, while  
we have investigated in \cite{SP98,ChPet991,ChPet992} 
the different physical problem
of {\it large} ``perturbations'' of density.
In \cite{Akh88} neutrino oscillations in a medium 
consisting of {\it even} number $n$
of alternating layers with densities 
$N_1$ and $N_2$ have been considered.
However,  i) the two layers were assumed to have 
equal widths, $L_1 = L_2$, and i) an enhancement of 
the neutrino transitions was 
found to take place for small vacuum mixing angles
at densities $N_{1,2}$ 
much smaller than the MSW resonance density.
The authors of \cite{Param86,Akh88} 
were interested in and found the conditions
for the classical parametric resonance 
in neutrino oscillations, which can take place 
after many periods of 
density modulation of the oscillations. 
From reading the articles \cite{Param86,Akh88,KS89}
it becomes clear that their authors had in mind
astrophysical applications of their results
\footnote{In all the concrete examples 
and corresponding plots 
in \cite{KS89}, for instance, the physical 
quantities having a dimension of length are 
given in units of the solar radius 
and are comparable with it; the values 
of the densities used 
in the examples differ substantially from 
those met in the Earth - they are noticeably 
smaller than the Earth mantle and core densities.}, and 
the results obtained in \cite{Param86,Akh88,KS89} 
may indeed have astrophysical applications.
Our studies \cite{SP98,ChPet991,ChPet992} 
were concerned primarily with the 
neutrino oscillations in the Earth. 
As a consequence,
the conditions of the enhancement 
of $P(\nu_a \rightarrow \nu_b)$
obtained in \cite{Param86,Akh88,KS89}
differ from those found 
by us in \cite{SP98,ChPet991,ChPet992}.
In the two- and three-layer medium cases 
i) and ii) discussed by us 
(e.g., in the case of the Earth)
the enhancement found 
in \cite{Akh88}, for example, is not realized.

 To summarize, after studying ref. \cite{ASNPB} 
and the references quoted therein
one can conclude  that the {\it new conditions for 
a total neutrino conversion in a medium} found and studied in
\cite{ChPet991,ChPet992} 
(i.e.,  the three sets of two conditions, eqs. 
(\ref{cond2L}), (\ref{cond3L}) and (\ref{cond2e3L}), and/or 
their solutions (e.g., eqs. (\ref{max-2A}) and (\ref{max-3A})), 
and/or and the regions where the solutions
can be realized, eqs. (\ref{region2A}) and (\ref{region3A})),
were indeed new: they 
were not derived and/or discussed in any form  
in \cite{Param86,Akh88,KS89}
or in any other article on the subject
of neutrino transitions in a medium
published before \cite{ChPet991,ChPet992}. 
None of them follows from the conditions
of enhancement of $P(\nu_a \rightarrow \nu_b)$
obtained in \cite{Param86,Akh88,KS89}
and thus they are not a particular case of the latter.
In \cite{ASNPB}, in particular,
the derivation of these conditions, first given 
in \cite{ChPet992}, is reproduced.
Most importantly, the {\it new 
conditions for a total neutrino conversion}
were shown in \cite{ChPet992} (see also \cite{SP98}) 
to be realized in the transitions in the Earth of the 
Earth-core-crossing  neutrinos (solar, atmospheric,
accelerator) and to lead to observable effects 
in these transitions - contrary to the claims 
made in \cite{ASNPB}.   
As for the physical interpretation of the
associated {\it new effect of total neutrino conversion}
in the cases of two-layer and three-layer
medium we have considered, we have proven
that this is a maximal constructive interference effect.
The interpretation of the {\it effect}
based on the expression 
(\ref{PAS}) for the neutrino transitions probabilities,
offered in \cite{ASNPB}, as we have pointed
out, is not convincing: expression (4), in particular,
has an artificial resonance-like form.
For the studies of the {\it new effect of total neutrino conversion}
in the Earth, eqs. (\ref{PCP2L}) and (\ref{PCP3L}) 
represent one of the several possible natural expressions 
for the relevant neutrino transition probabilities.
The rest is terminology.

\newpage

\end{document}